\documentclass[%
 reprint,
 superscriptaddress,
 amsmath,amssymb,
 aps,
 prc,
]{revtex4-2}

\newcommand*{\Ne}{\ensuremath{^{18}\textrm{Ne}(\alpha,p)^{21}\textrm{Na}}}
\newcommand*{\Ox}{\ensuremath{^{18}\textrm{O}(\alpha,p)^{21}\textrm{F}}} 
\newcommand*{\F}{\ensuremath{^{18}\textrm{F}(\alpha,p)^{21}\textrm{Ne}}}

\usepackage{graphicx}
\usepackage{dcolumn}
\usepackage{bm}
\usepackage{xcolor}

\begin{document}

\preprint{APS/123-QED}

\title{Direct measurement of $^{18}$F($\alpha,p$)$^{21}$Ne cross sections with ANASEN}

\author{K. S. Davis}
\email{Contact author: kdavis@triumf.ca}
\affiliation{Department of Physics \& Astronomy, Louisiana State University, Baton Rouge, LA 70803, USA}
\affiliation{TRIUMF, Vancouver, British Columbia V6T2A3, Canada}

\author{J. C. Blackmon}
\author{C. M. Deibel}
\author{G. L. Wilson}
\affiliation{
 Department of Physics \& Astronomy, Louisiana State University, Baton Rouge, LA 70803, USA}

\author{M. Alcorta}
\affiliation{TRIUMF, Vancouver, British Columbia V6T2A3, Canada}

\author{L. T. Baby}
\affiliation{Physics Department, Florida State University, Tallahassee, FL 32306, USA}

\author{D. W. Bardayan}
\affiliation{University of Notre Dame, Department of Physics, Notre Dame, Indiana 46556, USA}

\author{S. Carmichael}
\affiliation{University of Notre Dame, Department of Physics, Notre Dame, Indiana 46556, USA}

\author{S. Chakraborty}
\affiliation{School of Physics, Engineering and Technology,
University of York, Heslington, York, YO10 5DD, United Kingdom}
\affiliation{TRIUMF, Vancouver, British Columbia V6T2A3, Canada}

\author{C. Esparza}
\affiliation{Physics Department, Florida State University, Tallahassee, FL 32306, USA}

\author{J. Glorius}
\affiliation{GSI Helmholtzzentrum f\"ur Schwerionenforschung GmbH, Darmstadt, Germany}

\author{A. I. Karakas}
\affiliation{School of Physics \& Astronomy, Monash University, Clayton VIC 3800, Australia}

\author{A. Lennarz}
\affiliation{TRIUMF, Vancouver, British Columbia V6T2A3, Canada}

\author{B. Kay}
\affiliation{Physics Division, Argonne National Laboratory, Lemont, IL 60439, USA}

\author{J. Henning}
\affiliation{
 Department of Physics \& Astronomy, Louisiana State University, Baton Rouge, LA 70803, USA}

\author{G. W. McCann}
\affiliation{Physics Department, Florida State University, Tallahassee, FL 32306, USA}

\author{S. Pain}
\affiliation{Oak Ridge National Laboratory, Oak Ridge, TN 37830, USA}

\author{C. Ruiz}
\affiliation{TRIUMF, Vancouver, British Columbia V6T2A3, Canada}

\author{R. Russell}
\affiliation{School of Mathematics \& Physics, University of Surrey, Guildford, Surrey GU2 7XH, United Kingdom}

\author{V. Sitaraman}
\affiliation{Physics Department, Florida State University, Tallahassee, FL 32306, USA}

\author{B. Sudarsan}
\affiliation{
 Department of Physics \& Astronomy, Louisiana State University, Baton Rouge, LA 70803, USA}

\author{I. Tolstukhin}
\affiliation{Physics Division, Argonne National Laboratory, Lemont, IL 60439, USA}

\author{L. Wagner}
\affiliation{TRIUMF, Vancouver, British Columbia V6T2A3, Canada}

\author{I. Wiedenh\"over}
\affiliation{Physics Department, Florida State University, Tallahassee, FL 32306, USA}

\date{\today}

\begin{abstract}
\begin{description}
\item[Background]
The {\F} reaction may impact Asymptotic Giant Branch nucleosynthesis and helium burning on accreting white dwarfs. This reaction has never been directly measured, and constraints from a previous time-inverse measurement leave large uncertainties in the reaction rate.
\item[Purpose]
We measured {\F} cross sections directly for the first time covering from $2 \leq E_{cm} \leq 4$ MeV.
Combining these results with previous work and comparisons to statistical model calculations results in a substantial improvement in uncertainties in the reaction rate.
\item[Methods]
Cross sections were measured in inverse kinematics at TRIUMF-ISAC using a radioactive $^{18}$F beam and ANASEN with an extended $^{4}$He gas target. Protons were detected in arrays of silicon-strip detectors with the measured trajectories and energies allowing reconstruction of the center-of-mass energy and final state in $^{21}$Ne populated. 
\item[Results]
We found the total cross section is in good agreement with statistical calculations, though population of the second excited state in $^{21}$Ne is greater than predicted.  This direct measurement is combined with the previous time-inverse measurement and a new reaction rate is calculated for 0.1--3 GK.
\item[Conclusions]
The uncertainties in the {\F} reaction rate have been reduced to a sufficient level to allow robust predictions from AGB models. The new recommended rate leads to a 45\% increase in $^{19}$F production compared to the previous rate.
\end{description}
\end{abstract}

\maketitle


\section{Introduction}

The {\F} reaction is important in a variety of astrophysical environments, including Asymptotic Giant Branch (AGB) stars and accreting white dwarfs (WD). AGB stars are a post main-sequence phase of stellar evolution that occurs for low- and intermediate-mass stars of $\sim$1--8 $M_{\odot}$ \cite{Herwig2005}, and they are the astrophysical site responsible for producing elements such as carbon, nitrogen, and lead in our Galaxy \cite{Kobayashi2020}. A core primarily composed of carbon and oxygen is surrounded by distinct He- and H-burning shells and a large, outer convective envelope. Thermal pulses drive He-shell flashes that reach temperatures up to $T \sim 0.3$ GK in the He-burning shell. Convection zones then mix 
newly synthesized reaction products with the envelope, and stellar winds expel envelope material that enriches the interstellar medium \cite{Herwig2005}.

AGB modeling is sensitive to the input physics, including convective and non-convective mixing processes, as well as nuclear reaction rates. One sensitivity study looked specifically at the impact of the {\F} reaction rate upper and lower limits \cite{Lee2009} on final abundances of 77 isotopes in the He-shell across nine AGB models. It was shown that {\F} can be a proton source for the $^{18}$O$(p,\alpha)^{15}$N$(\alpha,\gamma)^{19}$F reaction chain and significantly increase the abundances of $^{19}$F and $^{21}$Ne \cite{Karakas2008}. Calculations with the upper limit rate, which is about 2 orders of magnitude larger than the lower limit at $T \sim 0.3$ GK \cite{Lee2009}, can also reproduce the high ratios of $^{21}$Ne/$^{22}$Ne observed in stellar silicon carbide grains \cite{Karakas2008}, the micron-sized, solid grains formed in AGB star envelopes that have been carried to Earth trapped in meteorites. Those abundances cannot, however, be reproduced when the recommended rate is used. This could suggest the {\F} reaction rate is closer to its upper limit than the recommended rate. However, the discrepancy between predicted and observed abundances could also be explained by the lack of well-understood non-convective mixing processes. 

The progenitor systems and mechanisms that lead to explosions of Type Ia supernovae (SNe Ia) remain open questions. The typical model includes a carbon-oxygen WD accreting material from a companion star in a binary system \cite{Hoyle1960, Iwamoto1999, Maoz2014}. The double-detonation theory, where a He flash on the surface of the WD generates a shockwave that ignites carbon fusion in the core and leads to a SNe Ia event, is one proposed model of the explosion mechanism. This can occur in binary systems where a WD accretes He-rich material from the donor star. 

Some accreting WD envelopes will have non-negligible amounts of $^{14}$N from the donor star, since $^{14}$N is a waiting point in the CNO cycle. During pre-shock He-burning when temperatures are less than 1 GK, the $^{14}$N can capture $\alpha$ particles and produce $^{18}$F via the $^{14}$N$(\alpha,\gamma)^{18}$F reaction. The timescale for this phase of the He flash is much faster than time required for the $^{18}$F abundance to be depleted by $\beta^{+}$ decay or $\alpha$ capture \cite{Shen2009}. After the shockwave is initiated, temperatures can increase to up to 3 GK at which point the {\F} reaction proceeds quickly. This can be a proton source for the $^{12}$C$(p,\gamma)^{13}$N$(\alpha,p)^{16}$O reaction chain during detonation \cite{Shen2009}. This series of reactions proceeds much faster than $^{12}$C directly capturing $\alpha$ particles and produces the same end product of $^{16}$O. Initial calculations with these reaction chains and the {\F} rate from Ref. \cite{Karakas2008} showed this He-accreting WD scenario could explain rare, less luminous subtypes of SNe Ia, often dubbed ``.Ia'' or ``Iax" \cite{Shen2009}.

Other recent double-detonation modeling with full reaction networks further support the importance of the $^{12}$C$(p,\gamma)^{13}$N$(\alpha,p)^{16}$O reaction chain \cite{Shen2014, Townsley2019, Wong2023} showing it reduces the He lifetime by up to four orders of magnitude compared to the triple-$\alpha$ reaction at post-shock temperatures of $T>1$ GK \cite{Shen2014}. Additionally, one-dimensional and multi-dimensional, hydrodynamic simulations have shown that double detonation with this reaction chain included in the nuclear network can lead to normal SNe Ia \cite{Townsley2019, Wong2023}. 

The {\F} reaction rate is governed by the properties of states in the compound nucleus $^{22}$Na. Early studies of $^{22}$Na focused on the low-energy states well below the Gamow window. In the astrophysically relevant range $E_{x} \gtrsim 8.6$ MeV, most measurements populate yrast states \cite{Lu1969, GomezDelCampo1973, Kadota1986, Vermeer1989}, which are not populated by the {\F} reaction. Garrett \textit{et al.} \cite{Garrett1971} measured energy levels up to 10.1 MeV with the $^{20}$Ne$(^{3}$He,$p)$ reaction, while Hallock \textit{et al.} \cite{Hallock1978} measured energy levels up to 17.9 MeV with the $^{12}$C$(^{14}$N$,\alpha)$ reaction. The density of states in this region is high, and some states are likely unresolved. Additionally, there is no angular distribution data and, therefore, no assignments or constraints on spin-parities of these levels.

In addition to these transfer measurements, Lee \textit{et al.} \cite{Lee2009} measured the time-inverse reaction $^{21} \text{Ne} (p,\alpha) ^{18} \text{F}$ by activating $^{21}$Ne-implanted targets with a proton beam. The $\beta^+$ decays of the resulting $^{18}$F were counted via coincident 511-keV photons from $e^{+}e^{-}$ annihilation \cite{Lee2009}. The measurement covered energies corresponding to  alpha energies of 0.6--1.5 MeV in the center-of-mass frame, and a reaction rate over the temperature range 0.1--1 GK was calculated. However, the time-inverse reaction study only constrains reactions populating the ground state,
\ensuremath{^{18}\textrm{F}(\alpha,p)^{21}\textrm{Ne}_{gs}}, and the contribution from reactions populating excited states in $^{21}$Ne is uncertain and could be substantial. A direct measurement of {\F} would determine the contributions from $^{21}$Ne excited states. Additionally, measurements over broader ranges of energies are needed to cover the Gamow window for temperatures $\gtrsim 1$ GK reached in the double-detonation model of SNe Ia progenitors.

\section{Methods}

\begin{figure}
\includegraphics[width=\linewidth]{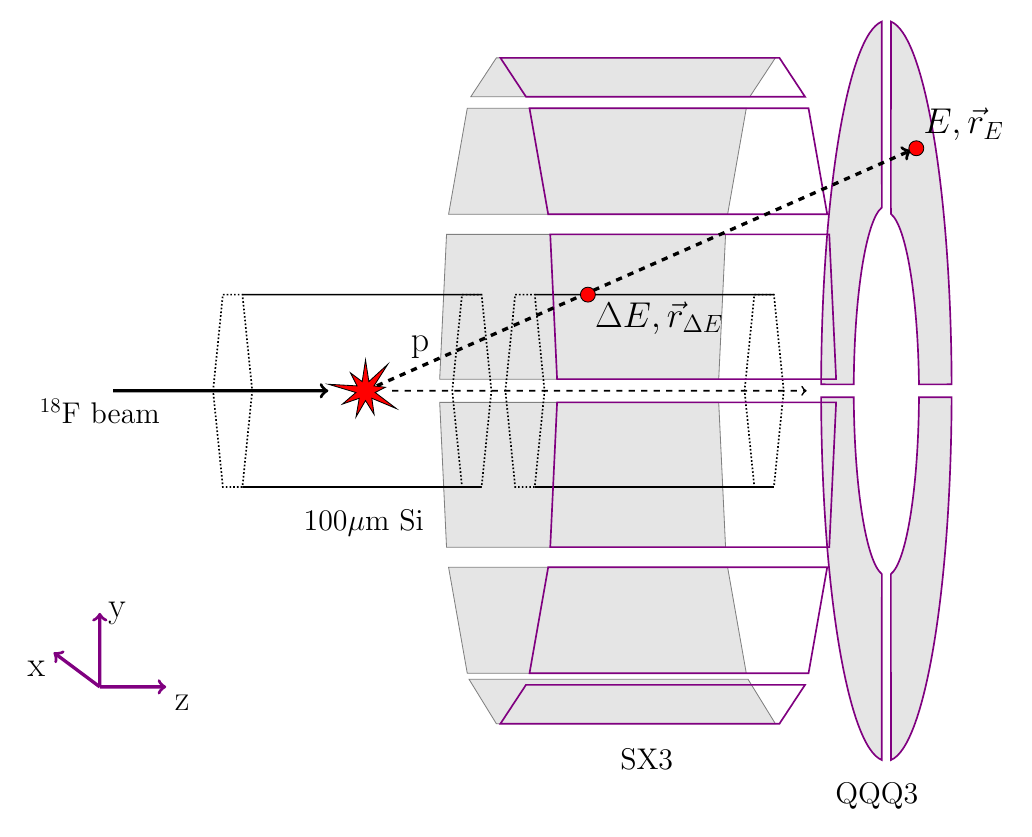}
\caption{\label{fig:geometry}A schematic of a {\F} reaction in the ANASEN chamber. The $^{18}$F beam enters the chamber through a window (not shown) and interacts with pure $^{4}$He gas. Protons at forward angles are detected by two sets of silicon arrays: the thin silicon detectors close to the beam axis measure energy loss, and the SX3 or QQQ3 detectors further from the beam axis measure residual energy.}
\end{figure}

We measured the {\F} excitation function in inverse kinematics at the TRIUMF ISAC-I facility 
using the Array for Nuclear Astrophysics and Structure with Exotic Nuclei (ANASEN), a gas target and charged-particle detector system designed for directly measuring $(\alpha,p)$ reactions \cite{Koshchiy2017}. An upgraded ANASEN design was used in this work as shown in Fig. \ref{fig:geometry}.

The beam enters the ANASEN chamber through a 3-$\mu$m-thick Mylar window into 175 Torr of pure $^{4}$He gas. The positions and energies of the outgoing protons are measured with a $\Delta E$ detector array close to the beam axis and a residual $E$ detector array farther from the beam axis. In this work, the $\Delta E$ array consists of two hexagonal barrels of 100-$\mu$m-thick silicon detectors (D+T Microelectronica CSIC-Barcelona). Each is segmented into 32 strips of 20-mm width and 2-mm pitch along the beam axis with a common back plane. The residual $E$ array includes a set of four 1000-$\mu$m-thick QQQ3-design detectors and a barrel of 12 1000-$\mu$m-thick Super X3-design (SX3) detectors (both Micron Semiconductor Ltd.) \cite{Koshchiy2017}. The position of the $\Delta E$ detector arrays relative to the $E$ detectors was chosen to optimize geometric coverage at more forward angles corresponding to higher proton energies.

Proton trajectories are  reconstructed from two measured positions on the silicon array and extrapolated back to the beam axis to determine the interaction point. Because the beam loses energy as it travels through the gas, the interaction point corresponds to the energy at which the reaction occurred, and the cross section can be measured at a range of reaction energies with one incident beam energy. 

Cross sections were measured using a 1.5-MeV/u $^{18}$F beam of overall intensity $10^5$--$10^6$ pps and 75\% purity. A 25-mm diameter, fast scintillator backed by a silicon photomultiplier array at zero degrees along the beam axis was used for beam normalization. A hybrid silicon ionization chamber (Si-IC) with $\sim$40 Torr of CF$_4$ gas was installed on a motion feedthrough in a vacuum chamber just upstream of the ANASEN chamber. It was inserted periodically throughout the experiment to monitor and optimize the beam composition, and retracted from the beam line during data collection. 
This work measured the {\F} excitation function at $E_{cm}$= 1.9--4 MeV simultaneously.

\section{Data Analysis}

\begin{figure}
\includegraphics[width=\linewidth]{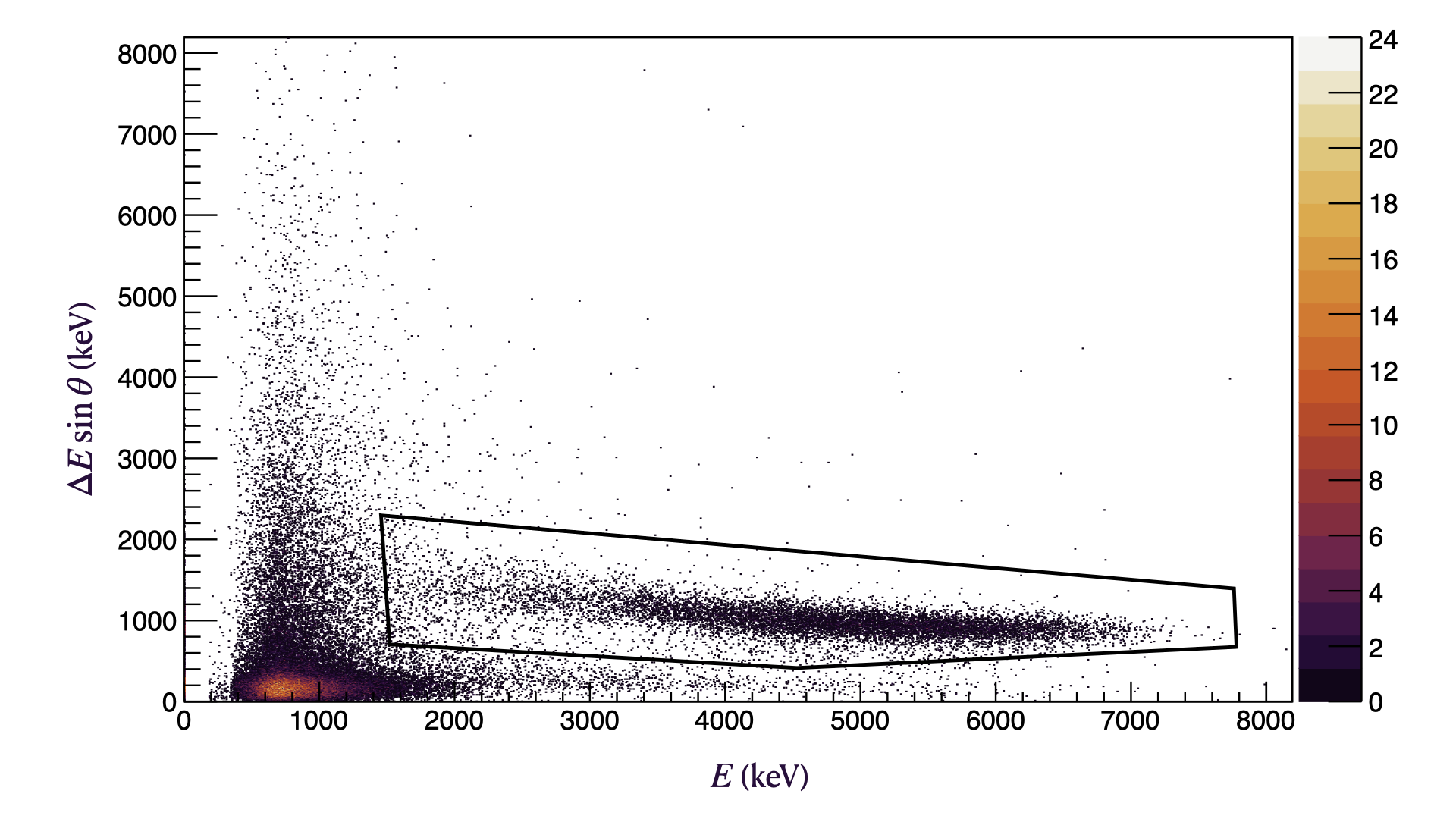}
\caption{\label{fig:particleID}Gain-matched, calibrated, and thickness-corrected $\Delta E$--$E$ particle identification plot, where $\Delta E$ is adjusted for detector thickness with a factor of $\sin \theta$. The proton gate is also shown in black.}
\end{figure}

Data was gain-matched and calibrated as described in Ref. \cite{Davis2026}. Protons are identified via $\Delta E$ signals from the inner detectors vs.\ residual $E$ signals from QQQ3 or SX3 detectors. The 
range of angles of incident protons results in a wide range of $\Delta E$ signals due to the range of effective silicon thickness. Particle identification resolution is improved by multiplying the energy loss  $\Delta E$ by $\sin \theta$ to correct for the effective thickness of silicon traversed. Figure \ref{fig:particleID} shows the proton gate applied to the angle-adjusted particle-identification plot. The proton group is cleanly identified, with $\Delta E \sin(\theta) \approx 1-1.5$ MeV as expected from stopping power estimates.

Reaction energies ($E_{cm}$) and total proton energies were calculated as detailed in Ref.\ \cite{Davis2026}. Briefly, the proton trajectory through the $\Delta E$ and $E$ detectors is extrapolated to the beam axis. The energy loss of the $^{18}$F beam through the window and corresponding length of gas to the reaction point is calculated using LISE++ \cite{Tarasov2008}, which determines the center-of-mass energy at which the reaction occured based only on the position measurements. The measured proton energies are corrected for energy loss in the gas to determine the proton energy immediately after the reaction and the excitation energy ($E_{f}$) of the final state populated in $^{21}$Ne.

\begin{figure}
\includegraphics[width=\linewidth]{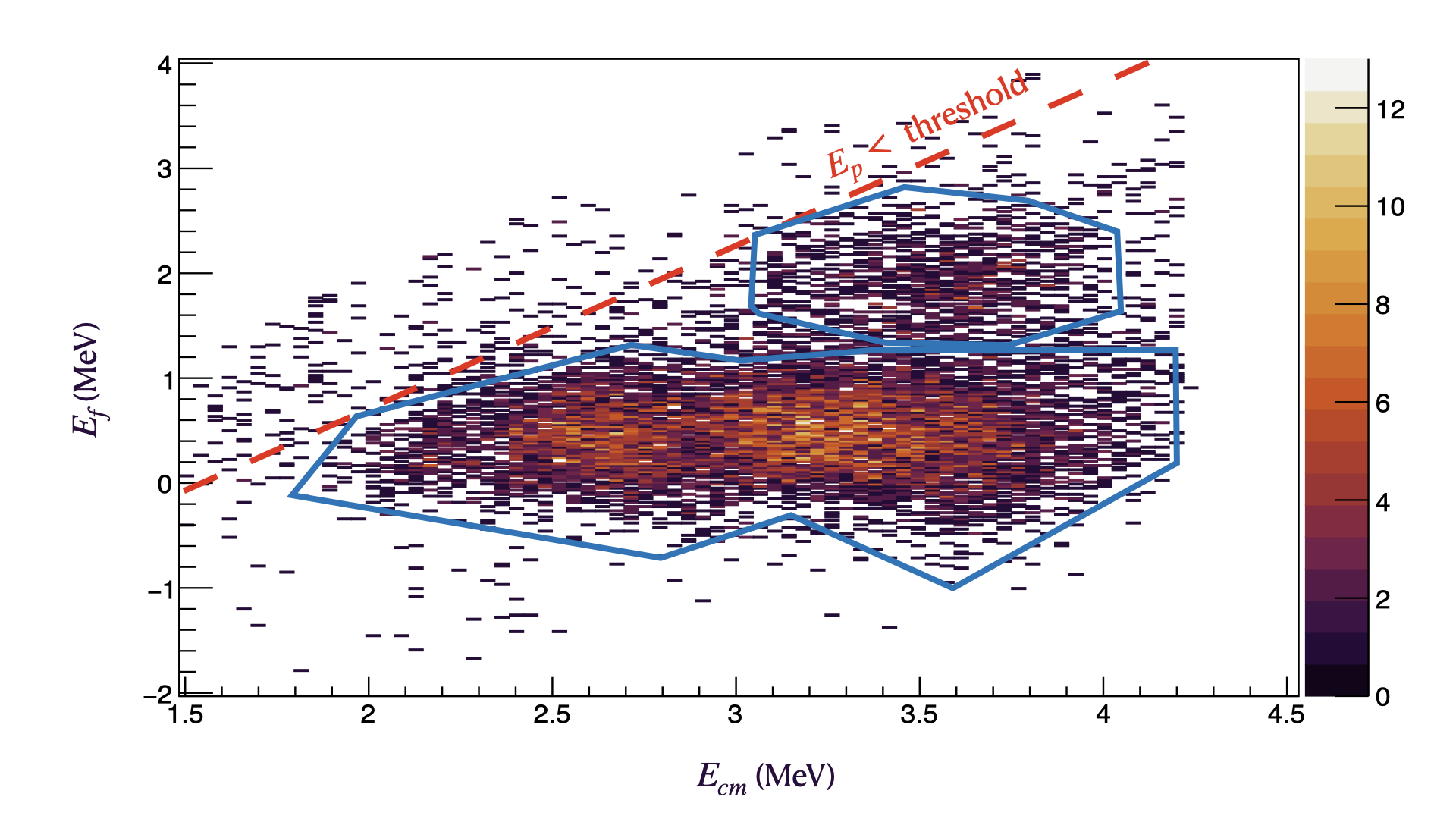}
\caption{\label{fig:ExEcm}A histogram of $^{21}$Ne excitation energy ($E_{f}$) vs. center-of-mass energy for {\F} events. The dashed red line depicts the energy threshold expected for a $\Delta E$--$E$ coincident signal. Gates for combined ground and 1st-excited states, and 2nd-excited state are shown in blue.}
\end{figure}

A histogram of $E_{f}$ versus $E_{cm}$ is shown in Fig. \ref{fig:ExEcm}. The dashed line shows the minimum detected proton energy required to generate $\Delta E$--$E$ coincident signals. The gap around $E_{cm}$ = 2.9 MeV is located at the physical gap between the two $\Delta E$ silicon arrays (see Fig. \ref{fig:geometry}). The ground state and 1st-excited state at 351 keV are observed for the full range of energies covered here, while the 2nd-excited state at 1746 keV is only observed at larger reaction energies of $E_{cm} \gtrsim 3$ MeV. 

\begin{figure}
\includegraphics[width=\linewidth]{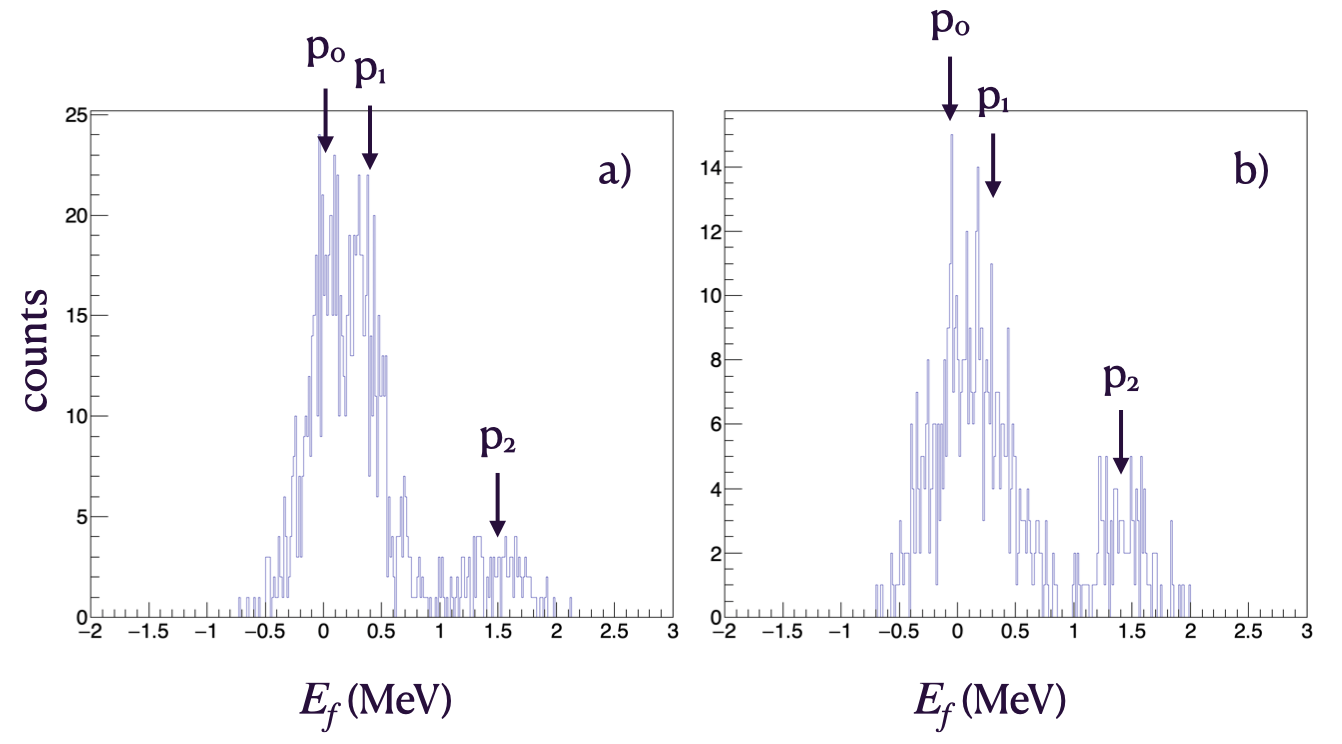}
\caption{\label{fig:Ex1D}Histograms of $E_{f}$ for (a) $E_{cm}$ = 3.3 MeV and (b) $E_{cm}$ = 3.4 MeV with ground, 1st-excited, and 2nd-excited states labeled $p_0$, $p_1$, and $p_2$, respectively.}
\end{figure}

Excitation-energy histograms gated on small ranges of $E_{cm} = 3.25$--3.30 MeV and 3.38--3.43 MeV are shown in Fig. \ref{fig:Ex1D}. The $p_0$, $p_1$, and $p_2$ labels indicate the expected $E_{f}$ values for reactions populating the ground, 1st-excited, or 2nd-excited states in $^{21}$Ne, respectively. As shown in Figs. \ref{fig:ExEcm} and \ref{fig:Ex1D}, the 2nd-excited state is resolved cleanly with a resolution of approximately 500-keV FWHM. However, the ground and 1st-excited states are not completely resolved in this work. Thus, the combined cross section from the ground and 1st-excited states are analyzed together. Only $E_{cm}$, which depends solely on geometry and tracking resolution, is used in further analysis.

\begin{figure}
\includegraphics[width=\linewidth]{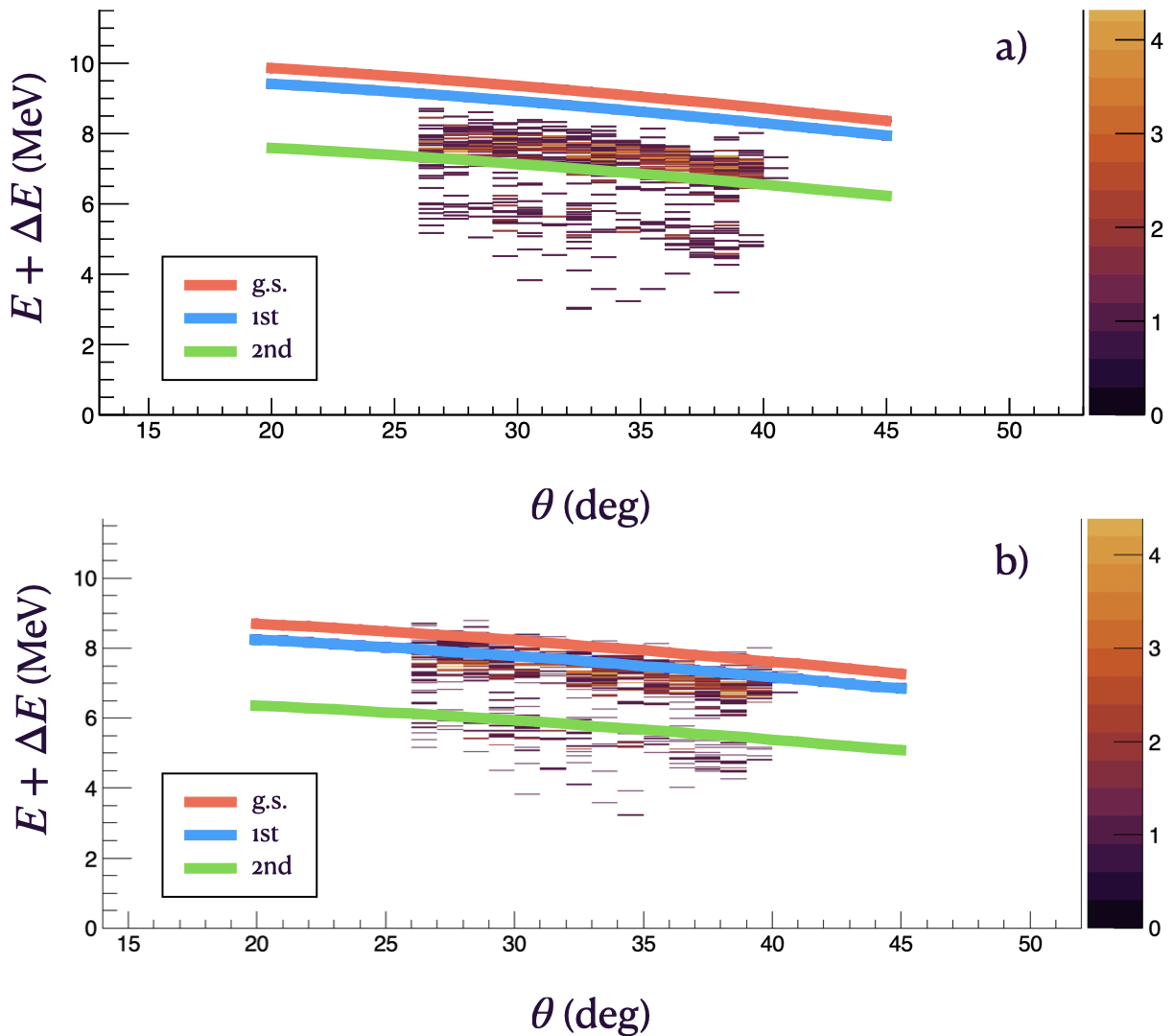}
\caption{\label{fig:noneon}Kinematic curves for protons from (a) {\Ne} and (b) {\F}. The expected energies and angles for $p_0$ (red), $p_1$ (blue), and $p_2$ (green) protons are overlayed on data from the {\F} reaction.}
\end{figure}

The beam contained an average of 2\% of $^{18}$Ne, which could contribute a proton background to the {\F} cross section. Kinematic curves for {\Ne} and {\F} were calculated using the relativistic kinematics code \textsc{Relkin} \cite{Drosg2017} and compared to the data. Figure \ref{fig:noneon} shows the expected angle and total energy for protons populating the first three states of the final nuclei at $E_{cm} = 3.3$ MeV with data gated on $E_{cm} =$ 3.28--3.32 MeV. As shown, the proton energies for {\Ne} are $\sim$1 MeV higher than those expected for {\F}. This is distinguishable with the demonstrated final state resolution $\delta E_{f} \lesssim 500$ keV. No protons are observed between 9--10 MeV as would be expected for reactions with $^{18}$Ne. Thus, we conclude background from the {\Ne} reaction is negligible. The remaining $^{18}$O in the beam does not produce any background due to the negative $Q$-value for the {\Ox} reaction. This was also confirmed during other runs with a pure, stable $^{18}$O beam.

\section{Results}
An experimental differential cross section was determined from the raw yield by
\begin{equation}
	\frac{d\sigma}{d\Omega} = \frac{Y}{I t \Delta\Omega \epsilon}
\end{equation}
where $Y$ is the measured proton yield, $I$ is the number of incident beam particles, $t$ is the target thickness in atoms/cm$^2$, $\Delta \Omega$ is the solid angle coverage of the detector, and $\epsilon$ is the detection efficiency. 
Data are divided into 0.5-cm bins along the beam axis, thus $Y$ is the yield per bin and the target thickness ($2.88\times 10^{18}$ atoms/cm$^2$) is calculated by multiplying the gas density by the bin size.

Solid angle coverage and detection efficiency were determined using a numerical Monte Carlo simulation \cite{Davis2026}. The simulation was also used to verify the 1--2 cm position resolution of the interaction point reconstruction, which depends on reaction position, and the proton energy and angle. For the $^{18}$F beam, this is equivalent to a 0.09--0.18 MeV reaction energy ($E_{cm}$) resolution in the center-of-mass frame.

\begin{figure}
\includegraphics[width=\linewidth]{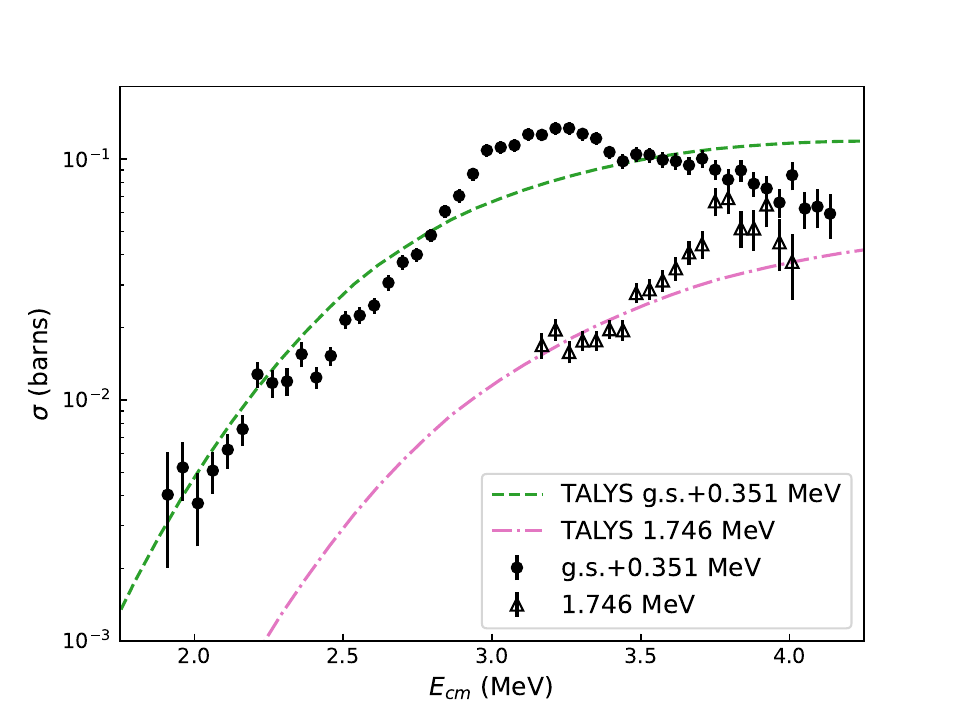}
\caption{\label{fig:datatalys}Combined cross section for the {\F} reaction populating the ground and 1st-excited states in $^{21}$Ne calculated with TALYS (dashed line) compared to this measurement (filled circles). TALYS calculated cross section (dot-dashed line) for the 2nd-excited state populated in $^{21}$Ne compared to this measurement (open triangles). Statistical error bars are shown. The TALYS cross sections were calculated with the alpha potential (``alphaomp 6'') from Ref. \cite{Avrigeanu2014}.}
\end{figure}

Cross sections from the data are compared to statistical model predictions from TALYS \cite{Koning2023} in Fig.\ \ref{fig:datatalys}. The default potential (labeled ``alphaomp 6'' \cite{Avrigeanu2014} in TALYS) was used for calculations in this work. The total measurement agrees within 30\% to TALYS predictions for the sum of reactions populating the ground, 1st-excited, and 2nd-excited states in $^{21}$Ne. It should be noted that the measured cross section for the 2nd-excited state is higher than the TALYS prediction by up to a factor of 2 at high energies ($E_{cm}$ $\geq$ 3.75 MeV) where the ground plus 1st-excited state measurement is lower than predicted by TALYS.

Due to the narrow center-of-mass angular coverage and limited angular distribution statistics for most reaction energies, we assume in calculating the total cross section that the average cross section over our measured angular range is the same as the average over all angles. To estimate the systematic uncertainty that results from our assumption of angular isotropy, we performed an $R$-matrix fit to the data using \textsc{azure2} \cite{Lane1958,Azuma2010}, and we generated 50 hypothetical angular distributions for cases that fit the data, and compared the average cross section over our measured angles to the average over all angles. We found there is a one-sigma variation in the average of --41\% to +31\%, which we adopt as the systematic uncertainty in our total cross section determination.

The TALYS {\F} predicted cross sections are particularly sensitive to the choice of $\alpha$-optical model parameters used. The largest cross sections from ``alphaomp 7'' \cite{Nolte1987} and ``alphaomp 8'' \cite{Avrigeanu1994} are inconsistent with our data and conservative systematic uncertainty, and are thus excluded from further analysis.
The ratios of cross sections from the five other optical models to the default model vary with energy and range between 0.6 and 2.

\section{Conclusions}

\begin{figure}
\includegraphics[width=\linewidth]{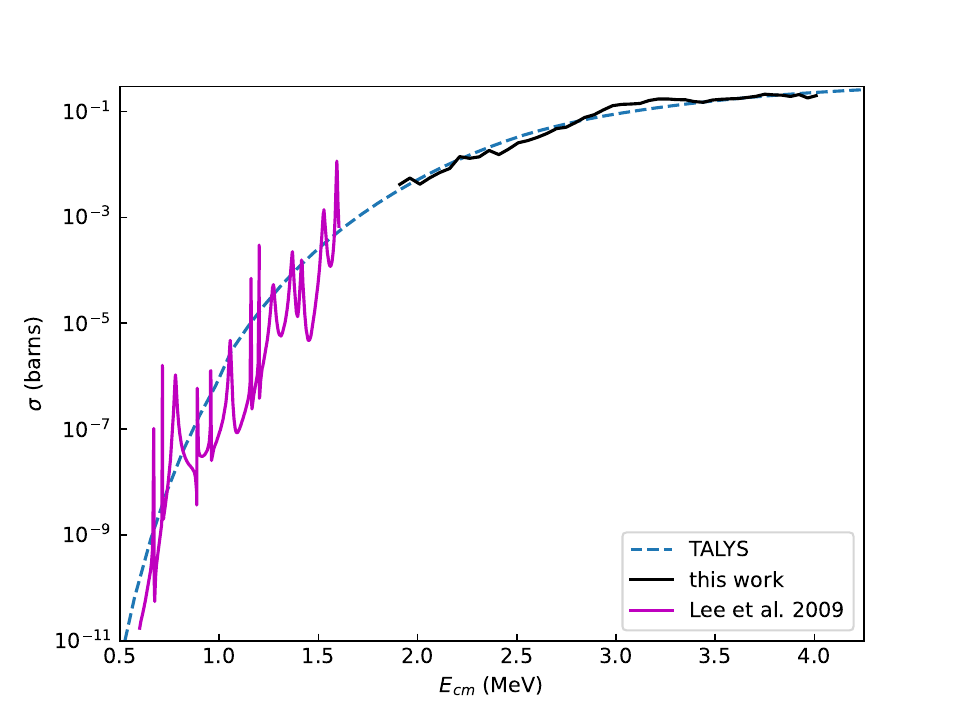}
\caption{\label{fig:xs}Total cross sections used in the reaction rate calculation. The corrected cross section from this work is in black, the corrected measurement from Ref. \cite{Lee2009} is in purple, and the Hauser--Feshbach cross section calculated with TALYS using ``alphaomp 6'' \cite{Avrigeanu2014} is in blue. The TALYS cross section is used for $E_{cm}$ = 0.1--0.6 MeV, 1.5--1.9 MeV, and 4.0--5.7 MeV.}
\end{figure}

The cross sections shown in Fig. \ref{fig:xs} were used to calculate the {\F} reaction rate using 
\begin{equation}
	N_A \langle \sigma v \rangle = \left(\frac{8}{\pi \mu}\right)^{1/2} \frac{N_A}{(k_B T)^{3/2}} \int_{0}^{\infty} E \sigma(E) e^{-E / k_B T} dE
\end{equation}
where $\mu$ is the reduced mass, $N_A$ is Avogadro's number, $k_B$ is the Boltzmann constant, and $T$ is the stellar temperature.

Comparisons between TALYS and measurements shows the ground state contributes 61--73\% of the total cross section for $0.6 < E_{cm} < 1.5$ MeV, and the ground state, first-excited, and second-excited states contribute 66--98\% of the total cross section for $1.9 < E_{cm} < 4.0$ MeV.  The measured cross sections were corrected for contributions from unobserved, higher lying final states by adding the TALYS predictions to the data. The corrected, total measured cross sections from Lee \textit{et al.} \cite{Lee2009} and this work were used from $0.6 < E_{cm} < 1.5$ MeV and $1.9 < E_{cm} < 4.0$ MeV, respectively. The TALYS total cross section was used for all other energies $E_{cm} \geq 0.1$ MeV. The resulting recommended rate for the temperature range 0.1--3 GK is shown in Table \ref{table:rate}.

The low rate was calculated using $- 1\sigma$ of the statistical and systematic uncertainty for the respective cross sections. This work has 4\%--15\% statistical uncertainty between $2.1 < E_{cm} < 4.0$ MeV, with up to 50\% at the lowest energies ($1.9 < E_{cm} < 2.1$ MeV), and --41\% to +31\% systematic uncertainty from the isotropic angular distribution assumption. The maximum 41\% uncertainty from Ref. \cite{Lee2009} is adopted for the cross section between $0.6 < E_{cm} < 1.5$ MeV. A factor of 0.6 is used for the TALYS cross section to account for uncertainty from choice of the $\alpha$-potential.

The high rate was calculated using $+ 1\sigma$ of the statistical and systematic uncertainty for the measured cross sections. The TALYS cross sections were replaced with the five non-excluded $\alpha$-potentials, and five separate rates were calculated at each temperature. The largest rate resulted from ``alphaomp3'' \cite{Demetriou2002} at low temperatures $T<0.8$ GK and ``alphaomp4'' \cite{Demetriou2002} at high temperatures $T>0.8$ GK, and was adopted as the high rate shown in Table \ref{table:rate}. The low rate and high rate are within a factor of $\sim$1.8--2.4 across all temperatures.

\begin{table}
\caption{\label{table:rate}Reaction rate $N_A \langle \sigma v \rangle$ of {\F} in cm$^3$ mol$^{-1}$ s$^{-1}$ units. The high rate and low rate are calculated using the $+1\sigma$ and $-1\sigma$ error bars, respectively, on the cross section.  See text for additional details.}
\begin{ruledtabular}
\begin{tabular}{ccccc}
Temp. & Recommended & Low & High & REACLIB \\
(GK) & & rate & rate & \\
\hline
0.1 & 2.90$\times$10$^{-22}$ & 1.91$\times$10$^{-22}$ & 3.47$\times$10$^{-22}$& 1.78$\times$10$^{-22}$ \\
0.15 & 2.67$\times$10$^{-17}$ & 1.86$\times$10$^{-17}$ & 3.33$\times$10$^{-17}$& 3.01$\times$10$^{-17}$ \\
0.2 & 3.45$\times$10$^{-14}$ & 2.50$\times$10$^{-14}$ & 4.45$\times$10$^{-14}$& 4.51$\times$10$^{-14}$ \\
0.25 & 6.90$\times$10$^{-12}$ & 4.75$\times$10$^{-12}$ & 9.32$\times$10$^{-12}$& 7.58$\times$10$^{-12}$ \\
0.3 & 6.31$\times$10$^{-10}$ & 3.94$\times$10$^{-10}$ & 8.79$\times$10$^{-10}$& 3.67$\times$10$^{-10}$ \\
0.4 & 3.78$\times$10$^{-7}$ & 2.25$\times$10$^{-7}$ & 5.33$\times$10$^{-7}$& 1.03$\times$10$^{-7}$ \\
0.5 & 1.97$\times$10$^{-5}$ & 1.16$\times$10$^{-5}$ & 2.77$\times$10$^{-5}$& 5.61$\times$10$^{-6}$ \\
0.6 & 2.79$\times$10$^{-4}$ & 1.65$\times$10$^{-4}$ & 3.94$\times$10$^{-4}$& 1.18$\times$10$^{-4}$ \\
0.7 & 1.96$\times$10$^{-3}$ & 1.15$\times$10$^{-3}$ & 2.76$\times$10$^{-3}$& 1.33$\times$10$^{-3}$ \\
0.8 & 9.70$\times$10$^{-3}$ & 5.74$\times$10$^{-3}$ & 1.36$\times$10$^{-2}$& 9.76$\times$10$^{-3}$ \\
0.9 & 4.24$\times$10$^{-2}$ & 2.53$\times$10$^{-2}$ & 5.88$\times$10$^{-2}$& 5.21$\times$10$^{-2}$ \\
1.0 & 1.74$\times$10$^{-1}$ & 1.05$\times$10$^{-1}$ & 2.38$\times$10$^{-1}$& 2.19$\times$10$^{-1}$ \\
1.5 & 3.51$\times$10$^{1}$ & 2.12$\times$10$^{1}$ & 4.64$\times$10$^{1}$& 3.14$\times$10$^{1}$ \\
2.0 & 7.72$\times$10$^{2}$ & 4.56$\times$10$^{2}$ & 1.02$\times$10$^{3}$& 6.31$\times$10$^{2}$ \\
2.5 & 6.10$\times$10$^{3}$ & 3.56$\times$10$^{3}$ & 8.08$\times$10$^{3}$& 4.83$\times$10$^{3}$ \\
3.0 & 2.78$\times$10$^{4}$ & 1.62$\times$10$^{4}$ & 3.68$\times$10$^{4}$& 2.12$\times$10$^{4}$ \\
\end{tabular}
\end{ruledtabular}
\end{table}

\begin{figure}
\includegraphics[width=\linewidth]{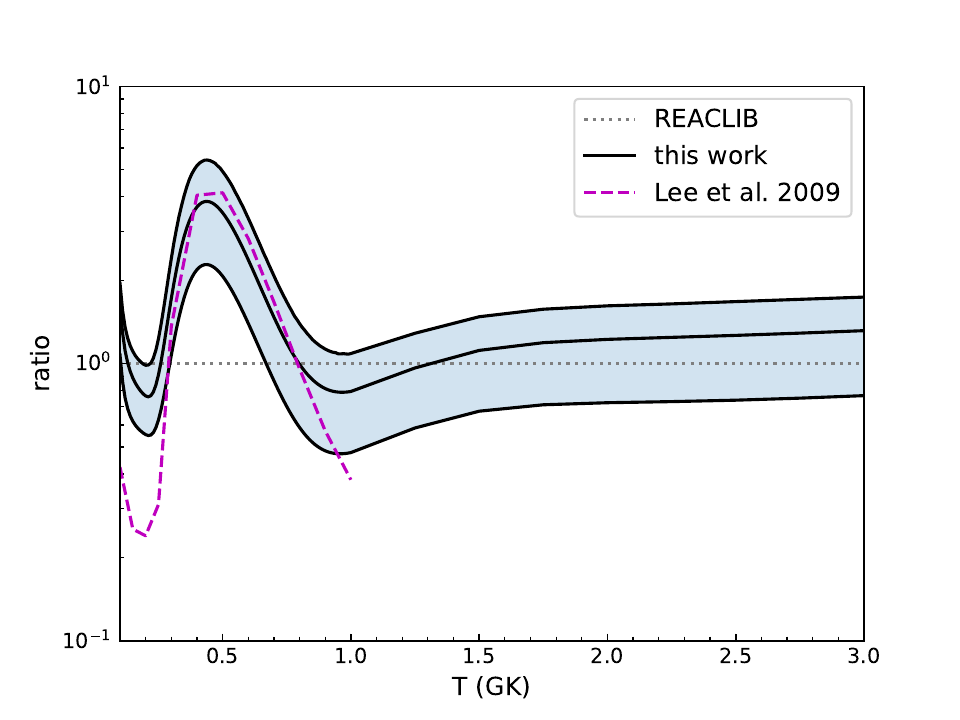}
\caption{\label{fig:ratioReaclib}Ratio of the new rate from this work and the previous rate from Ref. \cite{Lee2009} to the REACLIB rate over the temperature range 0.1--3 GK. Note that the rate in Ref. \cite{Lee2009} is only reported up to 1 GK. The low rate in this work is within a factor of $\sim$1.8--2.4 of the high rate.}
\end{figure}

The REACLIB \cite{Cyburt2010} recommended rate for {\F} is a theoretical rate calculated with the Hauser--Feshbach code \textsc{non-smoker} \cite{Rauscher1998}, and is in good agreement with the TALYS rate using the default $\alpha$-optical model potential. The ratios of rates from this work and Ref. \cite{Lee2009} to the REACLIB rate are shown in Fig. \ref{fig:ratioReaclib}. In the $T \lesssim 0.3$ GK region, the recommended rate is 0.7--1.2 times the REACLIB rate. In the temperature range $0.3 < T < 0.6$ GK, this rate is up to 4 times larger, with an upper limit up to 6 times larger. For $T \gtrsim 1$ GK, the REACLIB rate is smaller than the new rate but within the low and high rate uncertainty. The Lee \textit{et al.} \cite{Lee2009} rate is reported for 0.1--1 GK, and is up to 3 or 4 times lower than this work and REACLIB, respectively. We find good agreement between our measurement and the statistical model using the default $\alpha$-optical model potential and do not scale down the statistical cross section as was done in Ref. \cite{Lee2009}, which likely results in our higher recommended rate.

The impact of the new rate and reduced uncertainty on AGB models was studied using the stellar evolution models and numerical post-processing method described in Ref. \cite{Karakas2016}. The production of $^{19}$F in AGB stars peaks near 3 solar masses \cite{Lugaro2004}. Additional masses were included for reference and comparison to the previous rate \cite{Karakas2008}. All models use the mass loss rate from Ref. \cite{Vassiliadis1993}. A partial mixing zone (PMZ) is used to produce the $^{13}$C pocket, where the neutrons released by $^{13}$C($\alpha$,n)$^{16}$O can further increase $^{19}$F production \cite{Lugaro2012,Buntain2017}. The choice of PMZ size is further explained in Ref. \cite{Karakas2016}.

The results are summarized in Table \ref{table:agb} with the initial mass in solar units ($M_\odot$), initial metallicity ($Z$), and mass of the PMZ in solar units of each model. For each set of model parameters, the fluorine production is determined for our low, recommended, and high rates from Table \ref{table:rate}. The abundance $[\text{F}/\text{Fe}] = \log_{10} (N(\text{F})/(N(\text{Fe}))_{\text{surface}} - \log_{10} (N(\text{F})/N(\text{Fe}))_{\text{sun}}$ is the final surface abundance of fluorine relative to the sun. The yield $y$($^{19}$F) is the sum of fluorine produced and expelled in solar winds minus the initial fluorine present. Finally, the production factor $f(^{19}\text{F}) = \log_{10}(X_{\text{avg}}/X_0)$ is the ratio of average to initial fluorine mass fraction in the wind. The results from our recommended rate are within 10\% or less of results from our low and high rates, indicating that our results have reduced uncertainties in the reaction rate to a level to allow robust modeling.

\begin{table}
\caption{\label{table:agb}Summary of AGB modeling results using the recommended reaction rates in Table \ref{table:rate}.}
\begin{ruledtabular}
\begin{tabular}{cccccc}
$M_\odot$ & $Z$ & PMZ & [F/Fe] & $y$($^{19}$F) & $f$($^{19}$F) \\
\hline
2 & 0.0001 & 0 & 3.56 & 1.10e-5 & 3.51 \\

2 & 0.0001 & 0.002 & 3.68 & 1.47e-5 & 3.63 \\

3 & 0.001 & 0.002 & 1.330 & 1.12e-6 & 1.29 \\

3 & 0.014 & 0 & 0.722 & 3.30e-6 & 0.668 \\

3 & 0.014 & 0.002 & 0.883 & 5.20e-6 & 0.829 \\

5 & 0.001 & 0 & -0.10 & -7.32e-8 & -0.507 \\

5 & 0.014 & 0 & -0.10 & -6.59e-7 & -0.23 \\
\end{tabular}
\end{ruledtabular}
\end{table}

As seen in Table \ref{table:agb}, fluorine production increases with decreasing metallicity for each mass. The production factor, yield, and surface abundances are each significantly larger for low mass models than the intermediate mass models. These relative trends are consistent with prior work. 

However, the recommended rate presented in this work increases the mass fraction of fluorine in the wind ($X_{\text{avg}}$) by up to 45\% compared to the previous rate \cite{Karakas2008,Lee2009}. Our larger rate at $T \lesssim 0.3$ GK, the peak He-burning temperature in AGB stars, ultimately leads to higher overall $^{19}$F production. While this rate is larger than the previous recommended rate, it is nowhere near the 2--3 orders of magnitude increase required to explain the observed $^{21}$Ne/$^{22}$Ne observed in stellar silicon carbide grains.

The REACLIB rate was used in Refs. \cite{Shen2014, Townsley2019, Wong2023} in models of helium envelope detonation on C/O white dwarfs. The primary role of the {\F} reaction in these models is providing a proton source for the $^{12}$C$(p,\gamma)^{13}$N$(\alpha,p)^{16}$O reaction chain \cite{Shen2009}, which burns $^{4}$He 3--4 orders of magnitude faster than the triple-$\alpha$ reaction at temperatures $\gtrsim 1$ GK \cite{Shen2014}. The impact of the new recommended rate at large temperatures is small, but may increase the supply of protons that can capture onto $^{12}$C, which will further increase the $^{4}$He burn rate from this reaction chain. This may increase the likelihood of a subset of thermonuclear supernovae originating from this scenario. However, the increased rate in the temperature range $0.3 < T < 0.6$ GK may affect the availability of $^{18}$F at the onset of thermonuclear runaway. A full network calculation should be completed to determine the overall effect of the new recommended rate on these types of supernovae.

In summary, the {\F} reaction has been measured directly with ANASEN in inverse kinematics and shows good agreement with statistical models. The new recommended reaction rate is larger than previous work and enhances the production of $^{19}$F in AGB models. The uncertainty in the low and high rates for $T=0.1$--3 GK has been reduced to an overall factor of $\sim$1.8--2.4. This removes the {\F} reaction as a source of uncertainty in AGB modeling and allows more robust constraints on other input physics to match observed isotopic ratios. Additional modeling is needed to assess the impact of the reduced uncertainty on WD double detonation models.

\begin{acknowledgments}
This work was partially supported by the U.S. Department of Energy, Office of Science, Office of Nuclear Physics under contract numbers DE-SC0026091 and DE-FG02-96ER40978, National Science Foundation Graduate Research Fellowship Program under Grant No. GR-00010333, and by the National Science Foundation under Grant No. PHY-2012522. TRIUMF receives funding via a contribution through the National Research Council Canada.  

\end{acknowledgments}

\bibliography{Dissertation}

\end{document}